\documentclass[a4paper, amsfonts, amssymb, amsmath, reprint, showkeys, nofootinbib, twoside]{revtex4-1}
\usepackage[english]{babel}
\usepackage[utf8]{inputenc}
\usepackage[colorinlistoftodos, color=green!40, prependcaption]{todonotes}
\usepackage{comment}
\usepackage{amsthm}
\usepackage{mathtools}
\usepackage{physics}
\usepackage{xcolor}
\usepackage{graphicx}
\usepackage[left=23mm,right=13mm,top=35mm,columnsep=15pt]{geometry} 
\usepackage{adjustbox}
\usepackage{placeins}
\usepackage[T1]{fontenc}
\usepackage{lipsum}
\usepackage{csquotes}
\usepackage[pdftex, pdftitle={Article}, pdfauthor={Author}]{hyperref} 
\begin{document}
\title{XENON1T Excess: Some Possible Backgrounds}

\author{Biplob Bhattacherjee}
    \email[Correspondence email address: ]{biplob@iisc.ac.in}
    \affiliation{Center for High Energy Physics, Indian Institute of Science, Bangalore}
    
\author{Rhitaja Sengupta}
    \email[Correspondence email address: ]{rhitaja@iisc.ac.in}
    \affiliation{Center for High Energy Physics, Indian Institute of Science, Bangalore}


\begin{abstract}
This work is a study of some possible background sources in the XENON1T environment which might affect the energy spectrum of electronic recoil events in the lower side and might contribute to the observed excess. We have identified some additional possible backgrounds, like $^{41}$Ca, $^{49}$V, $^{63}$Ni, $^{106}$Ru and $^{125}$Sb coming from cosmogenic production, where the former two emit monoenergetic $X$-rays and the latter three have $\beta$ decays, or isotopes, like $^{210}$Pb, from the decay chain of $^{222}$Rn emanated in liquid xenon from the materials, or isotopes, like $^{137}$Cs, produced due to neutron capture. We perform a $\chi^2$ fitting of the ER spectrum from these backgrounds along with tritium to the observed excess events by varying their individual rates to understand whether they can be present to contribute to the low energy excess or their presence is constrained from the data. We also study the possibility of simultaneous presence of more than one such backgrounds, and how this affects the rates required by individual backgrounds to explain the excess.  
\end{abstract}


\maketitle


The XENON1T collaboration has recently observed an excess of electronic recoil (ER) events in the low energy region between 1-7 keV \cite{Aprile:2020tmw}. The possible new physics explanations outlined in their work are: solar axions, enhanced neutrino magnetic moment\footnotemark[1], and bosonic dark matter. However, they have clearly stated that one cannot rule out the hypothesis of beta decay of tritium being the source for this excess since the concentration of tritium in xenon have not been directly measured as yet. 
Out of the former three new physics hypotheses, the results of the solar axions and neutrino magnetic moment explanations are in tension with astrophysical constraints from stellar cooling \cite{Giannotti:2017hny}. 
There have been many recent works to explain this excess with various different new physics models \cite{Kannike:2020agf,Takahashi:2020bpq,Alonso-Alvarez:2020cdv,Boehm:2020ltd,Fornal:2020npv,Su:2020zny,Bally:2020yid,Harigaya:2020ckz,Du:2020ybt,Choi:2020udy,Chen:2020gcl,AristizabalSierra:2020edu,Bell:2020bes,Paz:2020pbc,DiLuzio:2020jjp,Buch:2020mrg,Dey:2020sai,Cao:2020bwd,Khan:2020vaf,Nakayama:2020ikz,Primulando:2020rdk,Lee:2020wmh,Robinson:2020gfu,Jho:2020sku,Baryakhtar:2020rwy,An:2020bxd,Bramante:2020zos,Bloch:2020uzh,Budnik:2020nwz,Zu:2020idx,Lindner:2020kko,Chala:2020pbn,Gao:2020wer,DeRocco:2020xdt,Dent:2020jhf,McKeen:2020vpf}. 

In this work, we take a different approach by studying some other possible backgrounds and their contribution to the ER spectrum.
This approach of looking into the backgrounds has been explored much less after XENON1T's recent result.
However, as is evident from the history of particle physics, observation of an excess demands that we carefully study our backgrounds in more detail before looking for possible new physics explanations.
This work is aimed at examining the various sources of backgrounds which might be present in the XENON1T environment, including the tritium beta decay, and 
pointing out the ones which can affect the low energy region of the ER spectrum. Even if they do not affect the low energy region enough to explain the observed excess, still their presence might reduce the significance of the excess. We are mostly concerned with the following:

\begin{itemize}

\item Are there any isotopes coming from cosmogenic production, either having a continuous $\beta$ spectrum or monoenergetic lines, which can contribute to the electronic recoil background?

\item Are there any other isotopes in the $^{222}$Rn decay chain, which is the source of the dominant background ($^{214}$Pb) in XENON1T, that might explain the excess? 

\item Are there any other isotopes coming from the neutron activation of xenon contributing to the background which have not been considered yet?

\item How does the simultaneous presence of more than one of these additional backgrounds, like multiple peaks or multiple continuous spectra or a peak along with a continuous spectrum, affect the observed excess?

\item How the presence of other backgrounds together with $^3$H can reduce the latter's rate and can such a combination provide a better fit of the data?

\end{itemize}

We later discuss some important questions related to purification that need to be addressed in the light of additional background sources.
Therefore, it is crucial that the XENON1T collaboration examines whether such backgrounds can actually be present, with the exact details of their purification process.



\footnotetext{The enhanced neutrino magnetic moment explanation is based on Refs.\cite{Bell:2005kz,Bell:2006wi}.}

\textit{Isotopes produced from cosmogenic activation of xenon having a $\beta$ decay} $-$
The cosmogenic activation of xenon 
can produce a number of different isotopes \cite{Zhang:2016rlz}, tritium being one of them. 
These isotopes are dominantly produced before transporting xenon to the LNGS underground hall, since their production rates scale with the flux of cosmic rays, and the latter drops drastically inside the LNGS cavern. 
Therefore, only isotopes with large half lives can survive and be present during the data taking runs of the experiment. 
Unless these are removed to a large extent by purification, their decays or transitions will contribute as backgrounds.

The lists of various isotopes which can be produced from the cosmogenic activation of xenon and other detector materials, as given in \cite{Zhang:2016rlz} and \cite{Piastra:2017ksr}, are not completely overlapping. Therefore, it is essential to have an exhaustive list of all isotopes that can be produced cosmogenically and to examine which of these can be potential backgrounds to the ER spectrum.
We have used \texttt{ACTIVIA} \cite{Back:2007kk} to identify all possible cosmogenic products that can come from xenon.
We use this as our starting point, and trace all possible decay chains of around 200 isotopes. We are only interested in those decay chains which have at least one isotope with half life long enough for it to survive the time gap between placing the xenon underground and starting of data taking (around 1000 days). 
We first look at isotopes having $\beta$ decays and discuss those which have monoenergetic transitions later.

Apart from half life of the isotope, there is another important point which determines which isotopes can be a possible background affecting the low energy ER spectrum. If the daughter particle of the isotope is dominantly produced in an excited state where it can emit prompt photons, then these fast emissions can shift the effective energy to the higher side.
To illustrate this, let us look at the isotopes of xenon, $^{133}$Xe and $^{135}$Xe. Although they have quite short half lives of 5.25 days and 9.14 hours respectively, and will mostly decay before the start of data taking, they can also be produced with neutron activation inside the detector once the data taking starts. 
$^{133}$Xe decays dominantly (98.5\%) to an excited state of $^{133}$Cs ($80.99$ keV) which emits a prompt photon ($T_{1/2}=6.283$ ns), and therefore, this gives a continuous energy background starting near this value. This background is already taken into consideration by the XENON1T collaboration. Similarly, the dominant decay of $^{135}$Xe is to $^{135}$Cs which is at an excited state of $249.77$ keV (96\%), and the latter emits photon with a half life of 0.28 ns. Therefore, this also does not contribute to the low energy ER spectrum.

Again, if the isotope has a very long half life, like $^{135}$Cs with a half life of few million years or $^{129}$I whose half life is one order greater than the former, then we might expect very less number of decays of these isotopes within the span of the data taking runs (around 227 days). 

The cosmogenically produced isotope, $^{125}$Sb has a half life of 2.76 years, so it can be present during data taking as well as have some decays during the data taking. It has a $Q$-value of 766.7 keV, which is on the higher side.
Along the lines discussed just above, most of the $^{125}$Te daughters of $^{125}$Sb are in excited states which radiate a photon promptly. However, there is one metastable state, $^{125}$Te$^{\text{m}}$, with an energy of 144.77 keV which has a half-life of 57.4 days. This particular decay has a branching of 13.6\% and can, therefore, be a background to the ER events.
However, we also need to take into account the effect of the peak at 144.77 keV on the full data $-$ (1-210) keV, which will almost have the same rate as the $^{125}$Sb $\beta$-decay.

We find that there are many more isotopes, in the list that we get using \texttt{ACTIVIA}, whose $\beta$ decay can affect the ER spectrum. Most of them have $\beta$ decays with higher $Q$-values, like $^{93}$Zr (90.8 keV),
$^{79}$Se (150.6 keV),
$^{14}$C (156.476 keV),
$^{35}$S (167.32 keV)\footnote{
Isotopes with $Q$-value greater than 200 keV are: 
$^{32}$Si (227.2 keV),
$^{45}$Ca (259.7 keV),
$^{135}$Cs (268.9 keV),
$^{99}$Tc (297.5 keV),
$^{90}$Sr (545.9 keV),
$^{39}$Ar (565 keV),
$^{42}$Ar (599 keV),
$^{85}$Kr (687 keV),
$^{36}$Cl (709.53 keV),
}.
However, their presence might contribute to the overall normalisation of the spectrum in low energy region.
We also identify some isotopes, like $^{63}$Ni and $^{106}$Ru, which have smaller $Q$-values of 66.977 keV and 39.4 keV respectively, and can therefore, contribute to the low energy ER excess. These isotopes satisfy the condition of having daughters with delayed decays or transitions, and have half lives of 101.2 years and 371.8 days respectively.
The former is an order larger than the tritium half life of 12.3 years, so there might be fewer decay events during the data taking, while a relatively smaller half life of the latter might lower its abundance in the experiment. Also their production cross sections are respectively three and four orders smaller than tritium production rate. 
However, since the condensation and purification processes suppress tritium by a factor of $\sim10^7$ \cite{XENON_slides}, these three or four orders of difference in production rates won't matter much if the purification processes are not as effective for $^{63}$Ni and $^{106}$Ru as for tritium, and they can be present in trace amounts.
Even if these are present or decay in trace amounts, that will be important for the ER spectrum and its low energy excess. 


\textit{The $^{222}$Rn decay chain} $-$
The dominant source of background in the XENON1T experiment for the ER events is the $\beta$ decay of $^{214}$Pb, which is a part of the $^{222}$Rn decay chain. $^{222}$Rn comes from the $^{238}$U decay chain which is present in the detector materials, like, stainless steel and PTFE \cite{Aprile:2015uzo}. It is important due to its relatively higher half life of 3.8 days and also due to the ability of Rn to diffuse into liquid xenon (LXe). Due to the much shorter half life of $^{220}$Rn (55.6 s), it has much less probability to diffuse into the active LXe volume.

In the decay chain of $^{222}$Rn, $^{214}$Pb is the first daughter that has a $\beta$ decay. 
As per our previous discussion, for a nuclide to be a possible background it has to decay to a state which has a delayed transition. 
For $^{214}$Pb, which itself has a half life of 27 mins, there is a 11\% branching to the ground state of $^{214}$Bi, which has a half life of around 20 mins. Therefore, it is a dominant background. $^{214}$Bi also has a $\beta$ decay, however, its daughter $^{214}$Po $\alpha$ decays to $^{210}$Pb with a very small half life of 164.3 $\mu$s and this helps to remove this background by vetoing the multiple scatter signals \cite{Aprile:2017fhu}.

$^{210}$Pb also has a $\beta$ decay and decays to the ground state of $^{210}$Bi with 16\% probability, where the latter has a half life of 5 days. Therefore, this can also be a possible background for the ER events. It has a $Q$-value of 63.5 keV, which is quite smaller compared to the $Q$-value of $^{214}$Pb ($\sim$1 MeV). This implies that $^{210}$Pb will have a more enhanced $\beta$ spectrum in the low energy region than $^{214}$Pb and therefore, might contribute to the excess in low ER energies. It decays to an excited state of $^{210}$Bi (46.539 keV) with 84\% branching, which emits a prompt photon to de-excite and therefore, shifts the $\beta$ spectrum. Since this energy lies within the range of the full data, we need to consider this as well while calculating the $\chi^2$. The half life of $^{210}$Pb is 22.2 years, which is of similar order of the tritium half life (12.3 years).

\textit{Isotopes from neutron activation of xenon} $-$
In the list of backgrounds considered in \cite{Aprile:2020tmw}, three of the backgrounds ($^{133}$Xe, $^{125}$I and $^{131\text{m}}$Xe) are produced from the neutron activation of xenon during neutron calibrations. We have thoroughly gone through all such neutron activated isotopes that can be produced with the isotopes of xenon present in its natural composition, to identify any additional source of background. We found that $^{137}$Xe (from $^{136}$Xe, $\sim$8.86\%) $\beta$ decays to the ground state of $^{137}$Cs (67\% branching\footnote{Rest of the branching is to excited states which lie outside 1-210 keV.}) with a half-life of $\sim$4 mins and $Q$-value of 4162.2 keV, and $^{137}$Cs $\beta$ decays to $^{137}$Ba with a half-life of $\sim$30 years and $Q$-value of 1175.63 keV. The branching to decay to the ground state of $^{137}$Ba is $\sim$5.3\% which is a stable isotope, and the rest goes to the metastable state $^{137\text{m}}$Ba with an energy of 661.66 keV ($T_{1/2}\sim 2.5$ mins). Therefore, these can have a flat contribution to the background and can affect its overall normalisation. Also the decay product of $^{135}$Xe, $^{135}$Cs has a $\beta$ decay to the ground state of $^{135}$Ba (100\% branching), with a $Q$-value of 268.9 keV, however, with a very large half-life of $2.3\times10^6$ years, and therefore, its effect will be highly suppressed. Two of the neutron activated isotopes, $^{125}$Xe and $^{127}$Xe, decay to excited states of their daughters ($^{125}$I and $^{127}$I) where the daughters emit prompt photons with energies 188.4 keV (25.3\% branching) and 202.86 keV (53\%) respectively, both of which lie within 1-210 keV. Therefore, they won't contribute to lower energies and might be constrained by the high energy data. 

\textit{Monoenergetic sources of background} $-$
Another possible explanation for the low energy excess might be due to the presence of some background which has a monoenergetic ER spectrum, for example, from the monoenergetic X-rays or Auger electrons coming from electron capture of some isotopes.
These isotopes can come from cosmogenic production in xenon, like $^{41}$Ca and $^{49}$V (having production rates three orders smaller than tritium, and half lives $10^5$ years and 330 days respectively) which have X-ray lines around 3-4 keV or can have some other source. The possibility of this excess coming from the 2.8 keV X-ray line of $^{37}$Ar, which is a cosmogenic product of xenon and is also injected during a dedicated calibration campaign in the final months of XENON1T’s operation, has been studied in \cite{Szydagis:2020isq}.

Let us now discuss briefly the purification process, which will affect mostly the cosmogenically produced isotopes.
Zirconium-based hot getters are used for the xenon gas purification.  
These getters can absorb hydrogen, which can contain traces of tritium. 
Therefore, the concentration of tritium present in xenon will be dependent on the ability of the zirconium getters to absorb hydrogen. 
If the getters get saturated before the concentration of $H_{2}$ in xenon is brought down to few ppb, this might account for the amount of tritium required to explain the low energy ER excess as has been pointed out in \cite{Robinson:2020gfu}. 

The background rate estimation of $^{125}$Sb, assuming that the zirconium getters don't remove Sb significantly, is quite large than the measured low energy ER background rate at the LUX experiment as has been shown in \cite{Baudis:2015kqa,Piastra:2017ksr}. Therefore, we might conclude that the getters have quite high efficiency for absorbing Sb. 
However, we could not find a detailed study which quantifies this efficiency with both hydrogen and Sb present, and whether Sb saturates the getters before the latter can remove hydrogen from xenon, is an important question.
The effect of purification on other isotopes like $^{63}$Ni, $^{106}$Ru, $^{41}$Ca or $^{43}$V and, in turn, the effect of their presence on the purification process of hydrogen and Sb needs to be studied. The getter's ability to absorb each one of them mostly depends on the electronegativity of the element. The existence of other isotopes might affect the purification for tritium, and even if it doesn't and tritium is completely removed from xenon, then these other isotopes might still be present and can be potential backgrounds. A detailed study of the purification efficiencies for each of these isotopes in the presence of each other is, therefore, much needed.

For backgrounds having $\beta$ decays, 
we select three cosmogenically produced isotopes ($^{63}$Ni, $^{106}$Ru, $^{125}$Sb) and one coming from the $^{222}$Rn decay chain ($^{210}$Pb). We also study the 3.314 keV and 4.511 keV $X$-ray lines of $^{41}$Ca and $^{49}$V respectively, as monoenergetic sources of background. 
We perform a simple binned $\chi^2$ fitting of the observed data with them as additional possible backgrounds. 
For the error associated with data, we have used the numbers as extracted from the XENON1T result. 

\begin{figure}
\centering
\includegraphics[width=8.6cm]{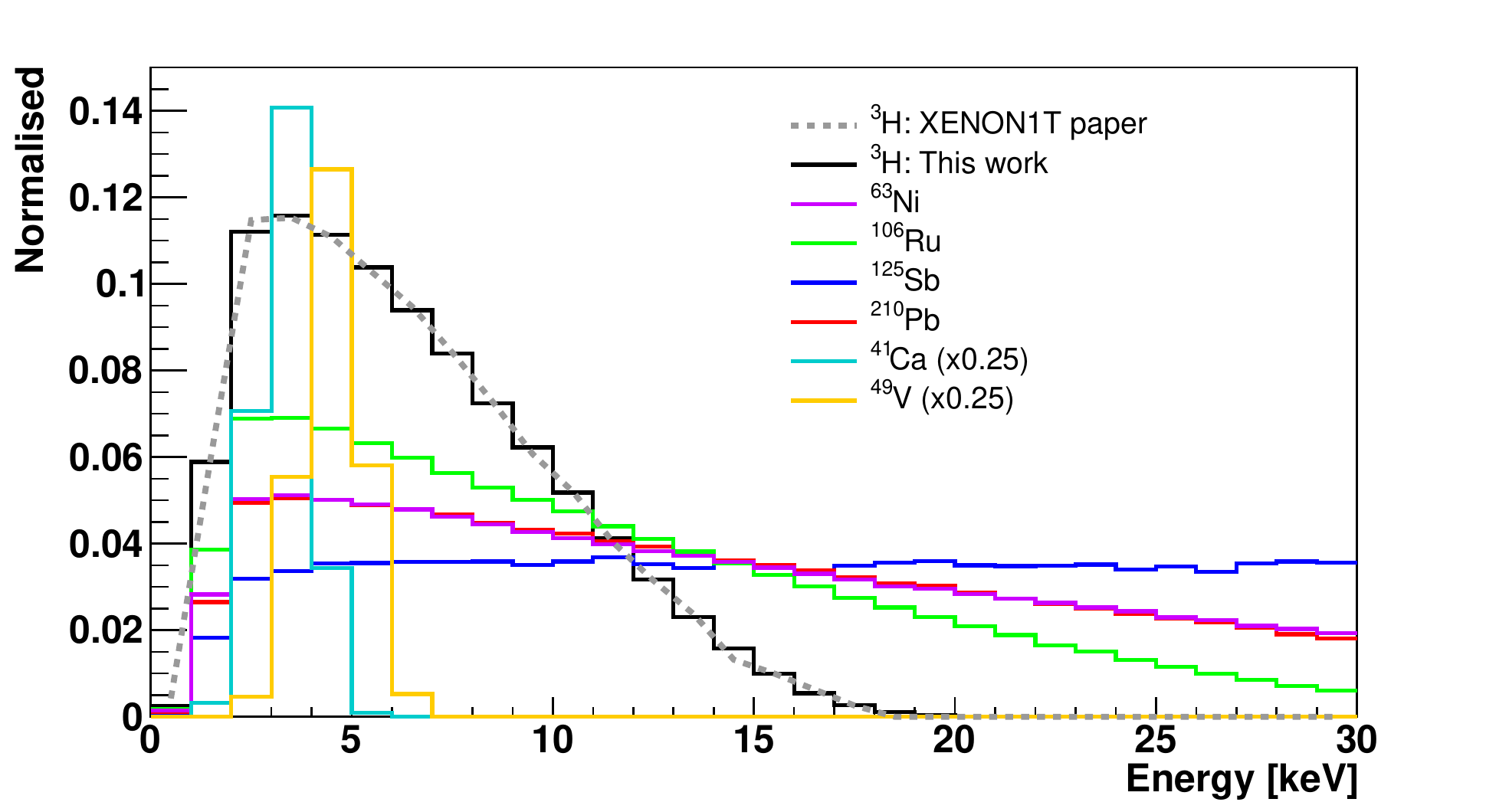}
\caption{Normalised distribution of energy from the $\beta$ decay of tritium ($^3$H), nickel ($^{63}$Ni), ruthenium ($^{106}$Ru), antimony ($^{125}$Sb), lead ($^{210}$Pb) and from monoenergetic $X$-rays of calcium ($^{41}$Ca) and vanadium ($^{49}$V), after applying the detector efficiency and resolution. The monoenergetic spectra are reduced in the plot by a factor of four. Also shown in gray is the distribution of $^3$H $\beta$ decay as obtained in \cite{Aprile:2020tmw}. The agreement in the shape validates our setup.}
\label{fig:beta_spectrum}
\end{figure}

\begin{figure*}
\centering
\includegraphics[width=\textwidth,height=4cm]{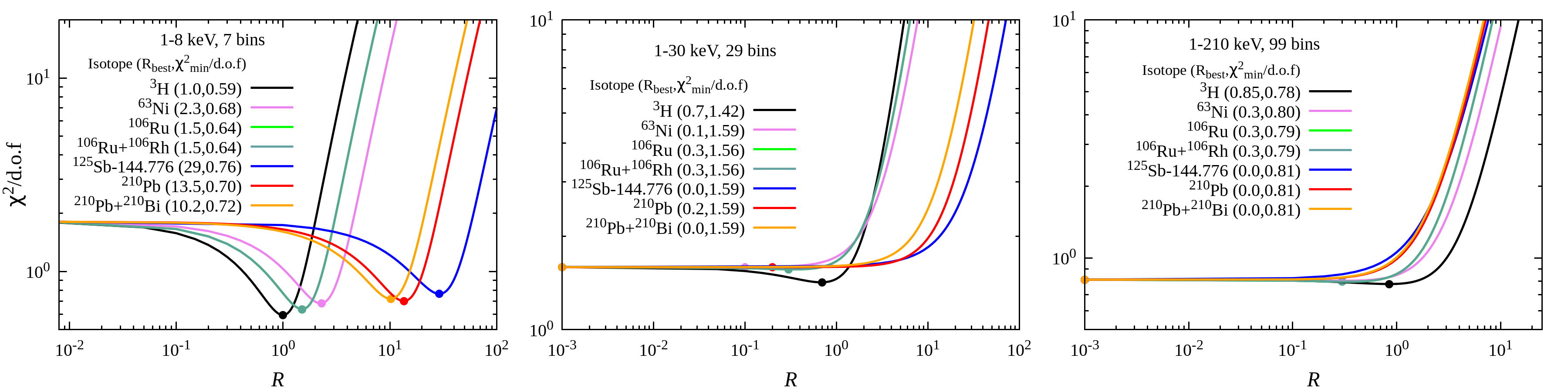}
\caption{$\chi^2$ fitting of the individual rates of tritium ($^3$H), nickel ($^{63}$Ni), ruthenium ($^{106}$Ru), antimony ($^{125}$Sb) and lead ($^{210}$Pb), along with all their associated energy spectra (like, $\beta$ spectrum of daughters $^{106}$Rh and $^{210}$Bi of $^{106}$Ru and $^{210}$Pb respectively), with varying factors of their rate with respect to the tritium rate fitted in \cite{Aprile:2020tmw} for energy range 1-8 keV with $N=7$ ({\it left}), 1-30 keV with $N=29$ ({\it centre}) and 1-210 keV with $N=99$ ({\it right}).}
\label{fig:1D_chi2}
\end{figure*}

Before proceeding with the $\chi^2$ fitting, we apply the detection and selection efficiencies as well as the energy resolution of the detector on the $\beta$ spectrum of each of the backgrounds. As a validation to this, we reproduce the energy spectrum coming from the tritium $\beta$ decay as given in \cite{Aprile:2020tmw}. Fig.\ref{fig:beta_spectrum} shows that these two (the solid black line and the gray dashed line) match very well and our setup is therefore validated. We now apply these efficiencies and resolutions on the 
energy spectra coming from the additional background sources that we have considered.
We take the $\beta$ spectra corresponding to the energy level of the daughter particles which does not have a very small half life ($<\mathcal{O}(1)$ ns) for reasons discussed in the previous sections. For $^{125}$Sb, it is the channel where $^{125}$Te$^{\text{m}}$ is the daughter particle with energy of 144.77 keV and for $^{210}$Pb, it is one where the daughter $^{210}$Bi is in the ground state. Both $^{63}$Ni and $^{106}$Ru decay to the ground state of their daughters with 100\% probability. 
We have used the $\beta$ spectrum as given in IAEA LiveChart (Nuclear Data Services database) \cite{NDS}, which they have obtained using BetaShape \cite{betashape} and \cite{PhysRevC.91.055504}.
The IAEA LiveChart (Nuclear Data Services database) calculations do not include exchange effects \cite{PhysRevC.91.055504} according to \cite{Aprile:2020tmw}, and therefore, they perform dedicated calculations to study the low energy discrepancies due to the exchange or screening effects for $^{214}$Pb and $^{85}$Kr. These effects might also be important for $^{125}$Sb and $^{210}$Pb and might affect their $\beta$ spectra at low energies. Such calculations are beyond the scope of this work, however, they must be included properly for a complete analysis. 
The $X$-ray energy values of $^{41}$Ca and $^{49}$V are also taken from the IAEA LiveChart database.
Fig.\ref{fig:beta_spectrum} shows the normalised energy distributions coming from the $\beta$ decays of $^{63}$Ni, $^{106}$Ru, $^{125}$Sb and $^{210}$Pb along with that of tritium, and the emitted $X$-rays of $^{41}$Ca and $^{49}$V. 

In addition to the $\beta$ spectrum of the isotopes that we have considered till now, many other energy spectra will be present associated with these isotopes. We need to check how much they are constrained by the data, which will put an indirect constraint on the background sources that we are discussing. We describe them in the following list:

\begin{itemize}
    \item $^{106}$Ru: $\beta$ decay of daughter $^{106}$Rh ($T_{1/2}=371.8$ days, $Q$-value=3545 keV) to the ground state of $^{106}$Pd with a branching of 78.6\%
    \item $^{125}$Sb: deexcitation of the daughter $^{125{\rm m}}$Te by emitting a photon of 144.776 keV ($T_{1/2}=57.4$ days)
    \item $^{210}$Pb: $\beta$ decay to the excited state of daughter, and the prompt emission of a photon of 46.539 keV (84\% branching) to the ground state of the daughter $^{210}$Bi, and its subsequent $\beta$ decay ($T_{1/2}\sim5$ days, $Q$-value=1161.2 keV) 
\end{itemize}
The $\beta$ decays of the daughter isotopes for $^{106}$Ru and $^{210}$Pb will affect the full energy range between 1-210 keV, however, those associated with the emission of prompt photons of energies 144.776 keV and 46.539 keV will have no effect on the data between 1-30 keV, and becomes important only after that. Since the 46.539 keV emission is prompt (3 ns) for the daughter $^{210}$Bi of $^{210}$Pb, it shifts the energy of $\beta$ decay of $^{210}$Pb to this excited state of the daughter. The resulting energy spectrum will be a sum of the monoenergetic 46.539 keV line and the original $\beta$ spectrum corresponding to this excited state.

We perform the $\chi^2$ fitting once with the first 7 bins since the observed excess is contained in this region, and then again over all the 29 bins corresponding to the energy range of 1-30 keV, and over 99 bins in the full 1-210 keV energy range. 
We take an agnostic approach and treat the rates of the additional backgrounds as floating parameters, perform the $\chi^2$ with varying rates of these over the standard backgrounds that have been considered in \cite{Aprile:2020tmw}, and find out the rate of each component corresponding to the minimum $\chi^2/\text{d.o.f}$. According to our new background hypothesis, the expected number of events in each bin is as follows:
\begin{equation}
n_{exp}^{i} = n_{B_0}^{i} + 
n_{\rm Additional~Background}^{i}\times R
\label{eq:bkg}
\end{equation}
where $n_{B_0}^{i}$ is the number of events in each bin $i$ coming from the backgrounds considered in \cite{Aprile:2020tmw} and $R$ is the rate of the additional background with respect to the tritium fitted rate of XENON1T, found out to be $159\pm51$ events/(t.y) \cite{Aprile:2020tmw}. We would like to remind here that these rates are the decay rates of the background isotopes and their actual concentration in xenon can be found out from these rates using their half lives and the branching to this particular decay channel.


Fig.\ref{fig:1D_chi2} shows the variation of the $\chi^2/\text{d.o.f}$ with varying rates of $^3$H, $^{63}$Ni, $^{106}$Ru, $^{125}$Sb and $^{210}$Pb w.r.t. the fitted tritium rate in \cite{Aprile:2020tmw} for energy range 1-8 keV with $N=7$ ({\it left}), 1-30 keV with $N=29$ ({\it centre}) and 1-210 keV with $N=99$ ({\it right}). The $\chi^2$/d.o.f values with only the $B_0$ hypothesis ($R=0$ for all additional backgrounds), as considered in \cite{Aprile:2020tmw}, are 1.81, 1.59 and 0.81 for energy range 1-8 keV with $N=7$, 1-30 keV with $N=29$ and 1-210 keV with $N=99$ respectively.
For tritium, the minimum $\chi^2/\text{d.o.f}$ is at $R=$1.0 and 0.7 when $N=7$ and 29 respectively, implying that the best fit rate for $N=7$ is equal to that found in \cite{Aprile:2020tmw}.

For $^{210}$Pb, the best fit rates from minimising the $\chi^2/\text{d.o.f}$ when $N=7$ and 29 are $R=$13.5 and 0.2 respectively - 16\% of which is actually contributing to the lower energies (84\% starts contributing only after 30 keV). Although the $\chi^2$ fitting over all 29 bins has minimum at a quite low rate of $^{210}$Pb, the best fit rate for $N=7$ lies within the $1\sigma$ interval of the $\chi^2$ fitting with $N=29$. The reduction in the tension with data due to the presence of $R=$13.5 of $^{210}$Pb is comparable to that of $^3$H. 
However, when we perform the $\chi^2$ on the full 1-210 keV data, the energy spectrum due to the dominant branching (84\%) of $^{210}$Pb to the excited state of $^{210}$Bi constraints the rate for $^{210}$Pb to about $\sim 2.4$ at 1$\sigma$. On checking the local significance, by requiring the $\chi^2$ of a single bin not to exceed 2.5$\sigma$ and not having two or more successive bins with $\chi^2$ > 2$\sigma$, the rate is further constrained to $R=0.6$. Such a rate will contribute to around 2 events in the 1-7 keV energy range in 0.65 tonne-year, which will have negligible effect in reducing the tension of the observed data. 
The addition of the $\beta$ decay of the daughter $^{210}$Bi (shown as the $^{210}$Pb+$^{210}$Bi line in fig.\ref{fig:1D_chi2}) does not affect the results much since it has a high $Q$-value and is mostly flat in the region between 1-210 keV.

Since $^{210}$Pb is a part of the $^{222}$Rn decay chain, its rate will be constrained by the observation of other decays that are also part of the same decay chain, like from the coincident $^{214}$Bi-Po and from the $\alpha$ decays of $^{218}$Po \cite{Aprile:2019dme}. In \cite{Aprile:2020tmw}, the fitted number of events of $^{214}$Pb $\beta$ decay, their dominant background source, was found to be 7480$\pm$160 in the 0.65 tonne-year exposure of SR1, which implies 11\,508$\pm$246 events/(t.y). Only 11\% of the $^{214}$Pb decays contribute as background to the experiment, therefore, there are around 104\,618$\pm$2236 $^{214}$Pb nuclei/(t.y) present in the detector. Since $^{214}$Pb, and its daughters $^{214}$Bi and $^{214}$Po, all have very small half-lives, they will all decay within the span of the experiment to give $^{210}$Pb. Owing to the large half-life of $^{210}$Pb of 22.2 years, it will have around 3216$\pm$69 $\beta$ decays/(t.y), 
corresponding to $R\approx$20.22, which lies just outside the $1\sigma$ interval of the $\chi^2$ fitting with $N=29$. 
This is a naive estimation and the actual rates will be reduced due to various detector effects, like plate-out. The strongest constraint on the rate of $^{210}$Pb, therefore, comes from the data.


For $^{125}$Sb, the best fit rates from minimising the $\chi^2/\text{d.o.f}$ when $N=7$ and 29 are $R=$29 and 0 respectively. 
The $N=29$ $\chi^2$, although having minimum at a very low rate can still allow larger rates till $\sim 20$ at 1$\sigma$.
However, in the full data till 210 keV, the monoenergetic peak of 144.776 keV coming from the $^{125{\text{m}}}$Te daughter will constrain the rates to $\sim 2$ at 1$\sigma$. Performing a check on the local significance, similar as the one described for $^{210}$Pb, the rate is further constrained to $R=0.2$. Therefore, larger rates of $^{125}$Sb which can reduce the tension with the data in low energies are not possible.
The presence of $^{125}$Sb can still play a pivotal role in affecting the purification of other species as we have previously discussed.

For $^{63}$Ni, the minimum $\chi^2/\text{d.o.f}$ is at $R=$2.3 and 0.1 when $N=7$ and 29 respectively, and for $^{106}$Ru, the minima are at $R=$1.5 and 0.3 for $N=7$ and 29 respectively. From the $\chi^2$ plots in the three energy ranges, we observe that a rate similar to the tritium rate will not affect the $\chi^2$ between 1-30 keV and 1-210 keV, however, it will help in reducing the tension in the low energy data, for both $^{63}$Ni and $^{106}$Ru.
For $^{106}$Ru, the $\beta$ spectrum of the daughter does not affect the results since it has a very high $Q$-value. Also, it has a higher half-life and therefore, its rate will be smaller compared to $^{106}$Ru.

Another cosmogenically produced isotope, $^{93}$Zr, has a $Q$-value of 90.8 keV and half-life of $1.61\times10^6$ years, which we have mentioned before. It has a 73\% branching to the 30.77 keV metastable excited state of $^{93\text{m1}}$Nb, where the latter has a half-life of 16.12 years. For this branching the $Q$-value of $^{93}$Zr will be reduced to $\sim$60 keV, and this can have some contribution in the low energy excess, similar to the isotopes of $^{63}$Ni and $^{106}$Ru. The large half-life of $^{93}$Zr imply that only few such $\beta$-decay events will be observed in the duration of the experiment, however, it might have the advantage of evading the purification which are performed using zirconium getters. Also, the rate of the peak at 30.77 keV will be further suppressed due to the large half-life, so we expect it to have little effect on the data.

Therefore, the $\chi^2$ study shows that although $^{210}$Pb and $^{125}$Sb can reduce the tension with the data in the low energy bins, their required rates are highly constrained from the full data till 210 keV. However we can have contributions from isotopes like $^{63}$Ni and $^{106}$Ru, which are not constrained by their associated spectra.




\begin{figure}[hbt!]
\centering
\includegraphics[width=8.6cm]{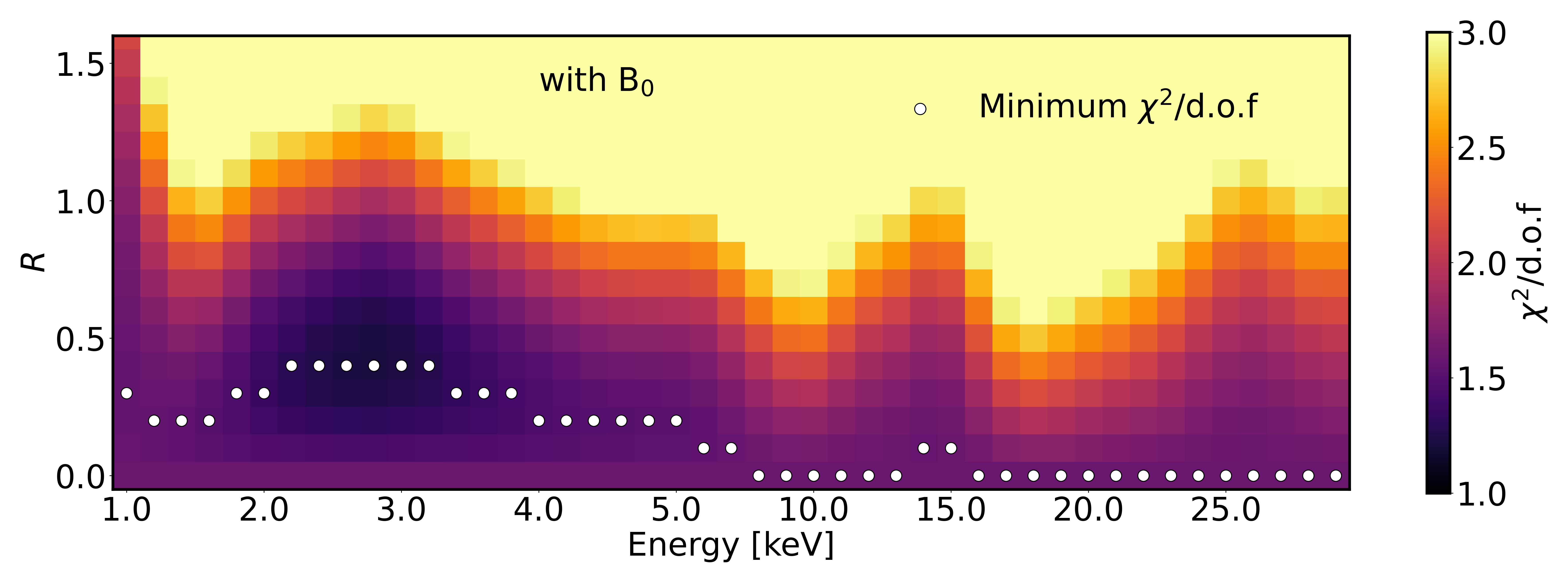}\\
\includegraphics[width=8.6cm]{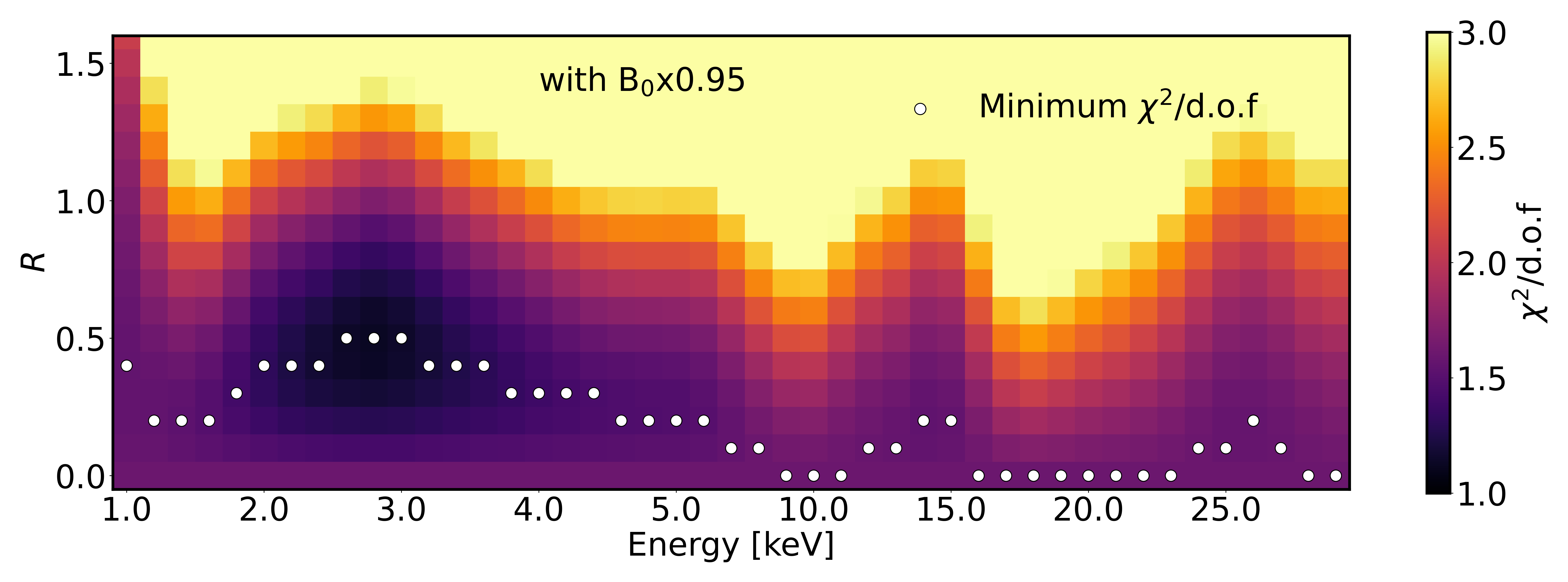}
\caption{$\chi^2$ fitting for 1-30 keV ($N=29$) of the observed data with monoenergetic peaks of different energies ({\it along X-axis}) added to the background in \cite{Aprile:2020tmw} ({\it top}) and with the latter reduced by 5\% ({\it bottom}), for varying rates $R$ ({\it along Y-axis}) with the $\chi^2$/d.o.f values shown in the color bar. The rates required for achieving the minimum $\chi^2$/d.o.f for each energy are shown as white circles. 
}
\label{fig:mono}
\end{figure}

We now discuss the fact that if this excess comes from some monoenergetic source, then which values of energy are favoured by the data. We have performed the $\chi^2$ analysis when monoenergetic peaks of different energies and intensities (or $R$) are added to the backgrounds considered in \cite{Aprile:2020tmw} and the result is shown in the {\it top panel} of fig.\ref{fig:mono}. In fig.\ref{fig:mono}, we have shown how varying the rates for each monoenergetic peak affects the $\chi^2$/d.o.f with the observed data for 1-30 keV. We have also marked the rates for each energy peak which achieves the minimum $\chi^2$ ({\it white circles}). For energies till 5 keV, we have performed a finer scan by varying the energy by 0.2 keV, and after that increase the energy of the peak by 1 keV till 29 keV.
We observe that monoenergetic lines with energies between $\sim2$ keV and $\sim4$ keV can reduce the tension with data. In addition, fig.\ref{fig:mono} conveys the fact that peaks with energies around 2-4 keV, 15 keV and 25 keV can be accommodated even with higher rates (around $R\sim1$) without increasing the $\chi^2$ much (within $1\sigma$).


We have redone the $\chi^2$ scan with monoenergetic peaks added with $B_0$, where the latter is reduced by 5\% and the result is shown in the {\it bottom panel} of fig.\ref{fig:mono}. We find that it can reduce the overall $\chi^2$ with the data than when using $B_0$ for many monoenergetic peaks, like between 2-4 keV, 14-15 keV and 25-26 keV. The dip in the data around 17 keV might also be an indication of a reduced baseline with multiple peaks present. 
The actual rate of the dominant background in the 1-30 keV range, $^{214}$Pb, might be smaller than the current fitted rate, and it can be compensated by other backgrounds like the $^{133}$Xe, which starts around $\sim$75 keV. As we are talking about the presence of multiple backgrounds, many of them can be simultaneously be present and affect the overall normalisation of $B_0$.

\begin{table}[hbt!]
    \centering
    \begin{tabular}{|c|c|c|c||}
    \hline
    Isotope & Energy [keV] & min($\chi^2$/d.o.f) & $R$ \\
    \hline\hline
    $^{37}$Ar & 2.8 & 1.21 & 0.41\\
    $^{41}$Ca & 3.314 & 1.31 & 0.4\\
    $^{49}$V & 4.511 & 1.50 & 0.2\\
    \hline\hline
    \end{tabular}
    \caption{Minimum $\chi^2$/d.o.f and $R$ values for X-rays coming from $^{37}$Ar, $^{41}$Ca and $^{49}$V added with the $B_0$ considered in \cite{Aprile:2020tmw} for $N=29$.}
    \label{tab:mono_isotope}
\end{table}

We have also shown the minimum $\chi^2$/d.o.f and $R$ values for X-rays of energy 2.8 keV, 3.314 keV and 4.511 keV coming from $^{37}$Ar, $^{41}$Ca and $^{49}$V respectively in table \ref{tab:mono_isotope}. Out of the many isotopes which can be produced cosmogenically, we have identified two ($^{41}$Ca and $^{49}$V) having $X$-rays in the keV energy range, however, there can be some others as well. Therefore, we have performed the scan shown in Fig.\ref{fig:mono} to visualise which monoenergetic peaks can reduce the tension with the data, and one can then identify isotopes which has $X$-ray lines corresponding to these energies.

One thing to note here is that the peak search performed in this work is different from the bosonic dark matter search presented in \cite{Aprile:2020tmw} in various aspects:\\

\begin{itemize}
\item The mass of the bosonic dark matter can be varied continuously, however, the $X$-ray lines coming from atomic transitions can have only some characteristic values. Although our result is shown as a continuous scan over the energy range, monoenergetic peaks can be obtained only at some specific values from the isotopes. We have performed the continuous search for illustration of how the $\chi^2$ would vary for different energies of the $X$-rays.
\item In our case, we can have more than one monoenergetic lines, for example, if both $^{41}$Ca and $^{49}$V are present and how such scenarios affect the fitting of the data will be shown in the following discussion. Also a single isotope can have multiple lines of various intensities, which can constrain their rates from the data. This is unlikely in the bosonic dark matter search where a peak is expected only at a given energy value depending on the dark matter mass.
\item If the monoenergetic peaks come from $X$-rays emitted by different isotopes, they can be present along with other isotopes having continuous $\beta$ energy spectrum. Together they can fit the data better as we have seen for $^3$H and $^{41}$Ca, where the former has a $\beta$ decay and the latter has an $X$-ray line at 3.314 keV. This kind of scenario is not considered in the bosonic dark matter search.
\end{itemize}

\begin{figure}[hbt!]
\centering
\includegraphics[width=7.cm,height=6.cm]{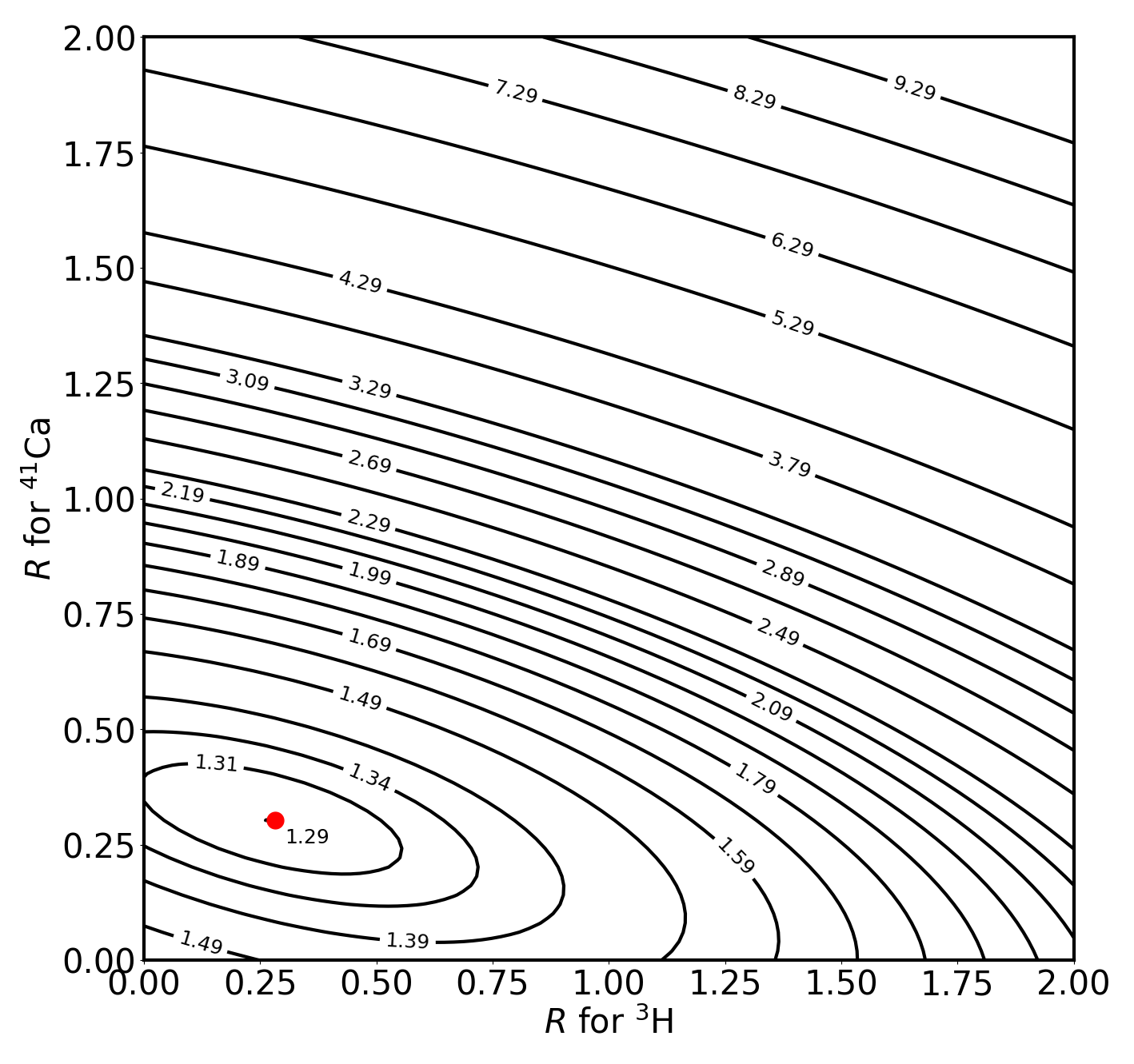}
\caption{Two dimensional $\chi^2$ contours for varying rates ($R$) of tritium ($^3$H) and calcium ($^{41}$Ca). The point corresponding to the best-fit rates is shown in {\it red}.}
\label{fig:2D_Ca_T}
\end{figure}

\begin{figure*}[hbt!]
\centering
\includegraphics[width=0.85\textwidth,height=4.5cm]{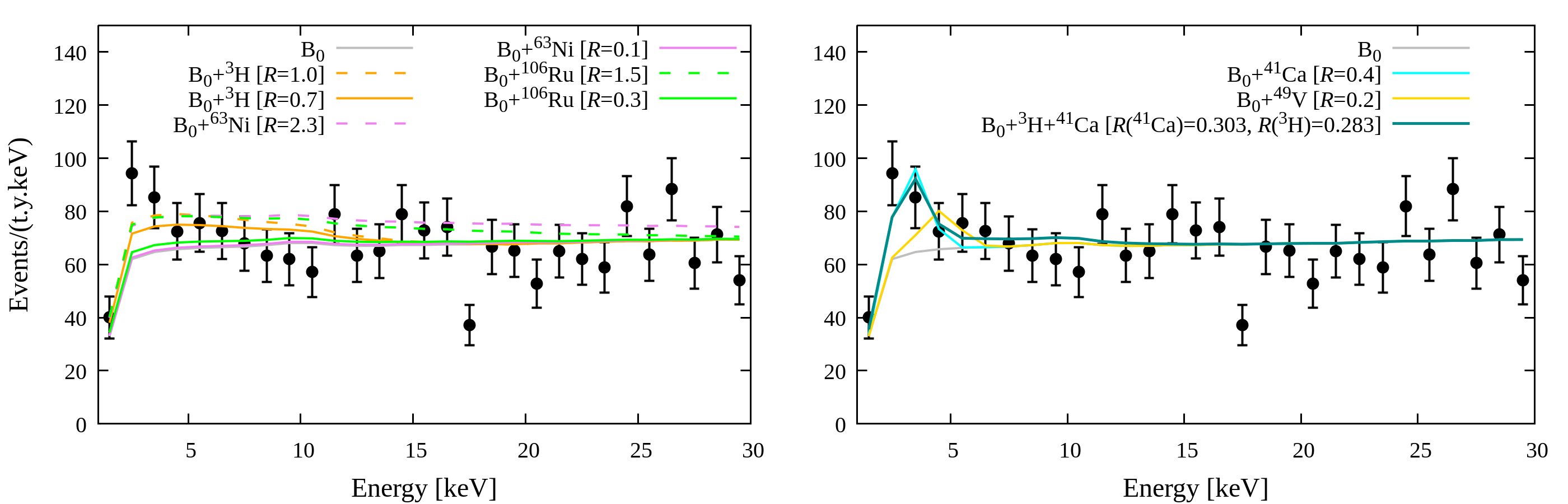}
\caption{Fits to the observed data with additional background hypotheses: $\beta$-decay spectra from tritium ($^3$H), nickel ($^{63}$Ni) and ruthenium ($^{106}$Ru) in the {\it left} plot; monoenergetic spectra from calcium ($^{41}$Ca) and vanadium ($^{49}$V) and simultaneous presence of monoenergetic line of $^{41}$Ca and $\beta$-decay spectrum of $^3$H in the {\it right} plot. The best-fit rates obtained from the $\chi^2$ fitting for 29 bins ({\it solid lines}) and 7 bins ({\it dashed lines}) are used.}
\label{fig:all_with_data}
\end{figure*}

For understanding the role of presence of multiple backgrounds on the observed excess, we perform two-dimensional $\chi^2$ fitting of the data with all possible combinations of the backgrounds that we have considered so far. We can have more than one monoenergetic sources, like $^{41}$Ca and $^{49}$V together, or a monoenergetic source along with a $\beta$ emitter, like $^{41}$Ca with $^3$H or $^{106}$Ru, or more than one $\beta$ emitters, like $^3$H and $^{106}$Ru or $^{106}$Ru and $^{63}$Ni. 
Fig. \ref{fig:2D_Ca_T} shows the two-dimensional $\chi^2$ contours with varying rates of $^3$H and $^{41}$Ca. The best fit rate giving the minimum $\chi^2/\text{d.o.f}$ value of 1.29 is 0.283 for $^3$H and 0.303 for $^{41}$Ca, and this is better than that achieved by both $^3$H and $^{41}$Ca individually. Also, even with a very low rate of $^3$H 
($R=0.1$), we can achieve similar quality of fit as achieved by $R=$0.7 of $^3$H individually in 1-30 keV region ($\chi^2/\text{d.o.f}=$1.42), if we add $^{41}$Ca with a rate of $R=0.1$. 
Similarly, a combination of $^{41}$Ca and $^{106}$Ru with $R=$0.162 and 0.465 respectively can also achieve a fit with $\chi^2/\text{d.o.f}\sim$1.42, which was achieved by $^3$H individually.
When two sources of monoenergetic lines, $^{41}$Ca and $^{49}$V are present together, they can achieve a fit with $\chi^2/\text{d.o.f}\sim$1.42 with rates of 0.222 and 0.283 respectively.

Considering pairs of continuous $\beta$ emitters, the central value of the best fit $\chi^2$ rate is driven by the isotope which fits the data better individually, however, non-zero rates of both the backgrounds can be accommodated within the data without affecting the fit much. For example, a combination of $R=0.1$ of $^3$H with $R=$0.869 of $^{106}$Ru can give a $\chi^2/\text{d.o.f}\sim$1.62, which is just 0.2 greater than that achieved by tritium alone. This combination shows that the tritium rate can be lowered by a factor of 7 (compared to $R=0.7$ for $N=29$) without affecting the quality of fit if another isotope, like $^{106}$Ru is also present. 
In addition, we also checked the fitting when more than two background sources are present together, and they can also be accommodated by the data. For example, when $^{63}$Ni, $^{106}$Ru and $^{41}$Ca are present in rates $R=$0.15, 0.1 and 0.15 respectively, they give $\chi^2/\text{d.o.f}\sim$1.42, similar to the $\chi^2$ value mentioned in the previous paragraph.
This exercise suggests that multiple backgrounds which were not considered previously can be present simultaneously and contribute to the XENON1T electronic recoil events.


For illustration of the fits of the various additional background hypotheses in this work with the observed data, we have shown in fig.\ref{fig:all_with_data} the XENON1T observed data in the 1-30 keV region along with the B$_{0}$ fit \cite{Aprile:2020tmw} and the backgrounds considered in this work. In the {\it left} plot, we show the $\beta$-decay spectra of tritium ($^3$H), nickel ($^{63}$Ni) and ruthenium ($^{106}$Ru) with the best fit rates obtained from the $\chi^2$ fitting for 29 bins ({\it solid lines}) and 7 bins ({\it dashed lines}). The {\it right} plot shows the fit to the observed data with monoenergetic spectra from calcium ($^{41}$Ca) and vanadium ($^{49}$V) and with the simultaneous presence of monoenergetic line of $^{41}$Ca and $\beta$-decay spectrum of $^3$H. We find that when both $^{41}$Ca and $^3$H are present, even for very small individual rates, together they can fit the data reasonably well. 

We have also performed all the above $\chi^2$ analyses with SR2 data which has been collected over a period of 24.4 days. The position of the $\chi^2$ minimum shifts slightly for some cases, however, the conclusions remain the same and the SR2 data can also not rule out the possible presence of these additional backgrounds. 
In summary, our endeavour here has been to perform an exhaustive search and examine some additional sources of backgrounds that might contribute to the XENON1T excess, unlike the many recent studies which attempt to explain the excess using various new physics models. We also want to highlight the possibility of multiple small backgrounds (multiple peaks or multiple continuous energy spectra, or both) contributing to the low energy excess. Through this study we arrive at the following conclusions:

\begin{itemize}

\item We have identified many different isotopes of cosmogenic origin having $\beta$ decays ($^{63}$Ni and $^{106}$Ru) or emitting monoenergetic $X$-rays ($^{41}$Ca and $^{49}$V) that might be potential backgrounds to the XENON1T experiment. All of these isotopes can reduce the tension of the observed excess in the low energy bins, comparable to that achieved by the tritium hypothesis without affecting the higher energy bins much. 
The rates required by $^{125}$Sb and $^{210}$Pb in the low energy region increases the $\chi^2$ with the overall data. The minimum value of $\chi^2$ fit with the $^{41}$Ca isotope is better than that with the $^3$H background in 1-30 keV energy range. 

\item The excess might be due to the presence of multiple lines, or even monoenergetic lines and $\beta$ emitters from many isotopes present together in small amounts, as has been shown by our two-dimensional $\chi^2$ fits. Presence of other isotopes, even in small rates can bring down the required $^3$H rate without affecting much the fit with the observed data (i.e., $\chi^2$ within $1\sigma$). The combination of $^3$H and $^{41}$Ca, where the latter has a monoenergetic $X$-ray spectrum, fits the data better than $^3$H and $^{41}$Ca individually, and at lower rates.

\item The fitted rate of $^{214}$Pb might be affected due to the presence of many other isotopes with high $Q$-values which have a flatter spectrum between 1-210 keV, like $^{137}$Cs coming from neutron activation, owing to the fact that the range of the expected rate of $^{214}$Pb is quite large ($\sim$ 5000 events in 0.65 tonne-year).

\item The fate of the cosmogenically produced isotopes depends on how the purification process works in removing each of them. Here one has to consider all of their presence to understand whether the getters get saturated before absorbing all of them completely. The presence of $^{125}$Sb might play an important role here, since they are absorbed by the getters and can affect the purification of other isotopes.


\end{itemize}

We have looked here only at the isotopes produced from cosmogenic activation of xenon and many other isotopes can be produced from the cosmogenic activation of the detector materials as well. These might also contain some potential backgrounds which have not yet been studied. Thus, a complete understanding of all such cosmogenic products is necessary. This work, therefore, emphasizes the need to study the backgrounds more carefully for XENON1T.

\textit{Acknowledgement} $-$
The work of B.B. is supported by the Department of Science and Technology, Government of India, under Grant No. IFA13- PH-75 (INSPIRE Faculty Award).

\appendix

\providecommand{\href}[2]{#2}\begingroup\raggedright\endgroup


\begin{thebibliography}{10}

\bibitem{Aprile:2020tmw}
{\bfseries XENON} Collaboration, E.~Aprile {\em et~al.}, ``{Observation of
  Excess Electronic Recoil Events in XENON1T},''
  \href{http://arxiv.org/abs/2006.09721}{{\ttfamily arXiv:2006.09721
  [hep-ex]}}.

\bibitem{Bell:2005kz}
N.~F. Bell, V.~Cirigliano, M.~J. Ramsey-Musolf, P.~Vogel, and M.~B. Wise,
  ``{How magnetic is the Dirac neutrino?},''
  \href{http://dx.doi.org/10.1103/PhysRevLett.95.151802}{{\em Phys. Rev. Lett.}
  {\bfseries 95} (2005) 151802},
  \href{http://arxiv.org/abs/hep-ph/0504134}{{\ttfamily arXiv:hep-ph/0504134}}.

\bibitem{Bell:2006wi}
N.~F. Bell, M.~Gorchtein, M.~J. Ramsey-Musolf, P.~Vogel, and P.~Wang, ``{Model
  independent bounds on magnetic moments of Majorana neutrinos},''
  \href{http://dx.doi.org/10.1016/j.physletb.2006.09.055}{{\em Phys. Lett. B}
  {\bfseries 642} (2006) 377--383},
  \href{http://arxiv.org/abs/hep-ph/0606248}{{\ttfamily arXiv:hep-ph/0606248}}.

\bibitem{Giannotti:2017hny}
M.~Giannotti, I.~G. Irastorza, J.~Redondo, A.~Ringwald, and K.~Saikawa,
  ``{Stellar Recipes for Axion Hunters},''
  \href{http://dx.doi.org/10.1088/1475-7516/2017/10/010}{{\em JCAP} {\bfseries
  10} (2017) 010}, \href{http://arxiv.org/abs/1708.02111}{{\ttfamily
  arXiv:1708.02111 [hep-ph]}}.

\bibitem{Kannike:2020agf}
K.~Kannike, M.~Raidal, H.~Veermäe, A.~Strumia, and D.~Teresi, ``{Dark Matter
  and the XENON1T electron recoil excess},''
  \href{http://arxiv.org/abs/2006.10735}{{\ttfamily arXiv:2006.10735
  [hep-ph]}}.

\bibitem{Takahashi:2020bpq}
F.~Takahashi, M.~Yamada, and W.~Yin, ``{XENON1T anomaly from anomaly-free ALP
  dark matter and its implications for stellar cooling anomaly},''
  \href{http://arxiv.org/abs/2006.10035}{{\ttfamily arXiv:2006.10035
  [hep-ph]}}.

\bibitem{Alonso-Alvarez:2020cdv}
G.~Alonso-Álvarez, F.~Ertas, J.~Jaeckel, F.~Kahlhoefer, and L.~Thormaehlen,
  ``{Hidden Photon Dark Matter in the Light of XENON1T and Stellar Cooling},''
  \href{http://arxiv.org/abs/2006.11243}{{\ttfamily arXiv:2006.11243
  [hep-ph]}}.

\bibitem{Boehm:2020ltd}
C.~Boehm, D.~G. Cerdeno, M.~Fairbairn, P.~A. Machado, and A.~C. Vincent,
  ``{Light new physics in XENON1T},''
  \href{http://arxiv.org/abs/2006.11250}{{\ttfamily arXiv:2006.11250
  [hep-ph]}}.

\bibitem{Fornal:2020npv}
B.~Fornal, P.~Sandick, J.~Shu, M.~Su, and Y.~Zhao, ``{Boosted Dark Matter
  Interpretation of the XENON1T Excess},''
  \href{http://arxiv.org/abs/2006.11264}{{\ttfamily arXiv:2006.11264
  [hep-ph]}}.

\bibitem{Su:2020zny}
L.~Su, W.~Wang, L.~Wu, J.~M. Yang, and B.~Zhu, ``{Xenon1T anomaly: Inelastic
  Cosmic Ray Boosted Dark Matter},''
  \href{http://arxiv.org/abs/2006.11837}{{\ttfamily arXiv:2006.11837
  [hep-ph]}}.

\bibitem{Bally:2020yid}
A.~Bally, S.~Jana, and A.~Trautner, ``{Neutrino self-interactions and XENON1T
  electron recoil excess},'' \href{http://arxiv.org/abs/2006.11919}{{\ttfamily
  arXiv:2006.11919 [hep-ph]}}.

\bibitem{Harigaya:2020ckz}
K.~Harigaya, Y.~Nakai, and M.~Suzuki, ``{Inelastic Dark Matter Electron
  Scattering and the XENON1T Excess},''
  \href{http://arxiv.org/abs/2006.11938}{{\ttfamily arXiv:2006.11938
  [hep-ph]}}.

\bibitem{Du:2020ybt}
M.~Du, J.~Liang, Z.~Liu, V.~Q. Tran, and Y.~Xue, ``{On-shell mediator dark
  matter models and the Xenon1T anomaly},''
  \href{http://arxiv.org/abs/2006.11949}{{\ttfamily arXiv:2006.11949
  [hep-ph]}}.

\bibitem{Choi:2020udy}
G.~Choi, M.~Suzuki, and T.~T. Yanagida, ``{XENON1T Anomaly and its Implication
  for Decaying Warm Dark Matter},''
  \href{http://arxiv.org/abs/2006.12348}{{\ttfamily arXiv:2006.12348
  [hep-ph]}}.

\bibitem{Chen:2020gcl}
Y.~Chen, J.~Shu, X.~Xue, G.~Yuan, and Q.~Yuan, ``{Sun Heated MeV-scale Dark
  Matter and the XENON1T Electron Recoil Excess},''
  \href{http://arxiv.org/abs/2006.12447}{{\ttfamily arXiv:2006.12447
  [hep-ph]}}.

\bibitem{AristizabalSierra:2020edu}
D.~Aristizabal~Sierra, V.~De~Romeri, L.~Flores, and D.~Papoulias, ``{Light
  vector mediators facing XENON1T data},''
  \href{http://arxiv.org/abs/2006.12457}{{\ttfamily arXiv:2006.12457
  [hep-ph]}}.

\bibitem{Bell:2020bes}
N.~F. Bell, J.~B. Dent, B.~Dutta, S.~Ghosh, J.~Kumar, and J.~L. Newstead,
  ``{Explaining the XENON1T excess with Luminous Dark Matter},''
  \href{http://arxiv.org/abs/2006.12461}{{\ttfamily arXiv:2006.12461
  [hep-ph]}}.

\bibitem{Paz:2020pbc}
G.~Paz, A.~A. Petrov, M.~Tammaro, and J.~Zupan, ``{Shining dark matter in
  Xenon1T},'' \href{http://arxiv.org/abs/2006.12462}{{\ttfamily
  arXiv:2006.12462 [hep-ph]}}.

\bibitem{DiLuzio:2020jjp}
L.~Di~Luzio, M.~Fedele, M.~Giannotti, F.~Mescia, and E.~Nardi, ``{Solar axions
  cannot explain the XENON1T excess},''
  \href{http://arxiv.org/abs/2006.12487}{{\ttfamily arXiv:2006.12487
  [hep-ph]}}.

\bibitem{Buch:2020mrg}
J.~Buch, M.~A. Buen-Abad, J.~Fan, and J.~S.~C. Leung, ``{Galactic Origin of
  Relativistic Bosons and XENON1T Excess},''
  \href{http://arxiv.org/abs/2006.12488}{{\ttfamily arXiv:2006.12488
  [hep-ph]}}.

\bibitem{Dey:2020sai}
U.~K. Dey, T.~N. Maity, and T.~S. Ray, ``{Prospects of Migdal Effect in the
  Explanation of XENON1T Electron Recoil Excess},''
  \href{http://arxiv.org/abs/2006.12529}{{\ttfamily arXiv:2006.12529
  [hep-ph]}}.

\bibitem{Cao:2020bwd}
Q.-H. Cao, R.~Ding, and Q.-F. Xiang, ``{Exploring for sub-MeV Boosted Dark
  Matter from Xenon Electron Direct Detection},''
  \href{http://arxiv.org/abs/2006.12767}{{\ttfamily arXiv:2006.12767
  [hep-ph]}}.

\bibitem{Khan:2020vaf}
A.~N. Khan, ``{Can nonstandard neutrino interactions explain the XENON1T
  spectral excess?},'' \href{http://arxiv.org/abs/2006.12887}{{\ttfamily
  arXiv:2006.12887 [hep-ph]}}.

\bibitem{Nakayama:2020ikz}
K.~Nakayama and Y.~Tang, ``{Gravitational Production of Hidden Photon Dark
  Matter in light of the XENON1T Excess},''
  \href{http://arxiv.org/abs/2006.13159}{{\ttfamily arXiv:2006.13159
  [hep-ph]}}.

\bibitem{Primulando:2020rdk}
R.~Primulando, J.~Julio, and P.~Uttayarat, ``{Collider Constraints on a Dark
  Matter Interpretation of the XENON1T Excess},''
  \href{http://arxiv.org/abs/2006.13161}{{\ttfamily arXiv:2006.13161
  [hep-ph]}}.

\bibitem{Lee:2020wmh}
H.~M. Lee, ``{Exothermic Dark Matter for XENON1T Excess},''
  \href{http://arxiv.org/abs/2006.13183}{{\ttfamily arXiv:2006.13183
  [hep-ph]}}.

\bibitem{Robinson:2020gfu}
A.~E. Robinson, ``{XENON1T observes tritium},''
  \href{http://arxiv.org/abs/2006.13278}{{\ttfamily arXiv:2006.13278
  [hep-ex]}}.

\bibitem{Jho:2020sku}
Y.~Jho, J.-C. Park, S.~C. Park, and P.-Y. Tseng, ``{Gauged Lepton Number and
  Cosmic-ray Boosted Dark Matter for the XENON1T Excess},''
  \href{http://arxiv.org/abs/2006.13910}{{\ttfamily arXiv:2006.13910
  [hep-ph]}}.

\bibitem{Baryakhtar:2020rwy}
M.~Baryakhtar, A.~Berlin, H.~Liu, and N.~Weiner, ``{Electromagnetic Signals of
  Inelastic Dark Matter Scattering},''
  \href{http://arxiv.org/abs/2006.13918}{{\ttfamily arXiv:2006.13918
  [hep-ph]}}.

\bibitem{An:2020bxd}
H.~An, M.~Pospelov, J.~Pradler, and A.~Ritz, ``{New limits on dark photons from
  solar emission and keV scale dark matter},''
  \href{http://arxiv.org/abs/2006.13929}{{\ttfamily arXiv:2006.13929
  [hep-ph]}}.

\bibitem{Bramante:2020zos}
J.~Bramante and N.~Song, ``{Electric But Not Eclectic: Thermal Relic Dark
  Matter for the XENON1T Excess},''
  \href{http://arxiv.org/abs/2006.14089}{{\ttfamily arXiv:2006.14089
  [hep-ph]}}.

\bibitem{Bloch:2020uzh}
I.~M. Bloch, A.~Caputo, R.~Essig, D.~Redigolo, M.~Sholapurkar, and T.~Volansky,
  ``{Exploring New Physics with O(keV) Electron Recoils in Direct Detection
  Experiments},'' \href{http://arxiv.org/abs/2006.14521}{{\ttfamily
  arXiv:2006.14521 [hep-ph]}}.

\bibitem{Budnik:2020nwz}
R.~Budnik, H.~Kim, O.~Matsedonskyi, G.~Perez, and Y.~Soreq, ``{Probing the
  relaxed relaxion and Higgs-portal with S1 \& S2},''
  \href{http://arxiv.org/abs/2006.14568}{{\ttfamily arXiv:2006.14568
  [hep-ph]}}.

\bibitem{Zu:2020idx}
L.~Zu, G.-W. Yuan, L.~Feng, and Y.-Z. Fan, ``{Mirror Dark Matter and Electronic
  Recoil Events in XENON1T},''
  \href{http://arxiv.org/abs/2006.14577}{{\ttfamily arXiv:2006.14577
  [hep-ph]}}.

\bibitem{Lindner:2020kko}
M.~Lindner, Y.~Mambrini, T.~B. de~Melo, and F.~S. Queiroz, ``{XENON1T Anomaly:
  A Light $Z^\prime$},'' \href{http://arxiv.org/abs/2006.14590}{{\ttfamily
  arXiv:2006.14590 [hep-ph]}}.

\bibitem{Chala:2020pbn}
M.~Chala and A.~Titov, ``{One-loop running of dimension-six Higgs-neutrino
  operators and implications of a large neutrino dipole moment},''
  \href{http://arxiv.org/abs/2006.14596}{{\ttfamily arXiv:2006.14596
  [hep-ph]}}.

\bibitem{Gao:2020wer}
C.~Gao, J.~Liu, L.-T. Wang, X.-P. Wang, W.~Xue, and Y.-M. Zhong,
  ``{Re-examining the Solar Axion Explanation for the XENON1T Excess},''
  \href{http://arxiv.org/abs/2006.14598}{{\ttfamily arXiv:2006.14598
  [hep-ph]}}.

\bibitem{DeRocco:2020xdt}
W.~DeRocco, P.~W. Graham, and S.~Rajendran, ``{Exploring the robustness of
  stellar cooling constraints on light particles},''
  \href{http://arxiv.org/abs/2006.15112}{{\ttfamily arXiv:2006.15112
  [hep-ph]}}.

\bibitem{Dent:2020jhf}
J.~B. Dent, B.~Dutta, J.~L. Newstead, and A.~Thompson, ``{Inverse Primakoff
  Scattering as a Probe of Solar Axions at Liquid Xenon Direct Detection
  Experiments},'' \href{http://arxiv.org/abs/2006.15118}{{\ttfamily
  arXiv:2006.15118 [hep-ph]}}.

\bibitem{McKeen:2020vpf}
D.~McKeen, M.~Pospelov, and N.~Raj, ``{Hydrogen portal to exotic
  radioactivity},'' \href{http://arxiv.org/abs/2006.15140}{{\ttfamily
  arXiv:2006.15140 [hep-ph]}}.

\bibitem{Zhang:2016rlz}
C.~Zhang, D.~M. Mei, V.~Kudryavtsev, and S.~Fiorucci, ``{Cosmogenic Activation
  of Materials Used in Rare Event Search Experiments},''
  \href{http://dx.doi.org/10.1016/j.astropartphys.2016.08.008}{{\em Astropart.
  Phys.} {\bfseries 84} (2016) 62--69},
  \href{http://arxiv.org/abs/1603.00098}{{\ttfamily arXiv:1603.00098
  [physics.ins-det]}}.

\bibitem{Piastra:2017ksr}
F.~Piastra, \href{http://dx.doi.org/10.5167/uzh-142441}{{\em {Materials
  radioassay for the XENON1T dark matter experiment, and development of a time
  projection chamber for the study of low-energy nuclear recoils in liquid
  xenon}}}.
\newblock PhD thesis, Zurich, U., 11, 2017.

\bibitem{Back:2007kk}
J.~Back and Y.~A. Ramachers, ``{ACTIVIA: Calculation of Isotope Production
  Cross-sections and Yields},''
  \href{http://dx.doi.org/10.1016/j.nima.2007.12.008}{{\em Nucl. Instrum. Meth.
  A} {\bfseries 586} (2008) 286--294},
  \href{http://arxiv.org/abs/0709.3472}{{\ttfamily arXiv:0709.3472 [nucl-ex]}}.

\bibitem{XENON_slides}
 E.~Shockley (for XENON Collaboration),
 ``{Search for New Physics with Electronic-Recoil Events in XENON1T},''
 \href{https://www.dropbox.com/s/vrq49tldjdfvnjh/shockley20200617.pdf?dl=0}{{\em LNGS Webinar, June 17th, 2020}}

\bibitem{Aprile:2015uzo}
{\bfseries XENON} Collaboration, E.~Aprile {\em et~al.}, ``{Physics reach of
  the XENON1T dark matter experiment},''
  \href{http://dx.doi.org/10.1088/1475-7516/2016/04/027}{{\em JCAP} {\bfseries
  04} (2016) 027}, \href{http://arxiv.org/abs/1512.07501}{{\ttfamily
  arXiv:1512.07501 [physics.ins-det]}}.

\bibitem{Aprile:2017fhu}
{\bfseries XENON} Collaboration, E.~Aprile {\em et~al.}, ``{Intrinsic
  backgrounds from Rn and Kr in the XENON100 experiment},''
  \href{http://dx.doi.org/10.1140/epjc/s10052-018-5565-y}{{\em Eur. Phys. J. C}
  {\bfseries 78} no.~2, (2018) 132},
  \href{http://arxiv.org/abs/1708.03617}{{\ttfamily arXiv:1708.03617
  [astro-ph.IM]}}.
  
\bibitem{Szydagis:2020isq}
M.~Szydagis, C.~Levy, G.~Blockinger, A.~Kamaha, N.~Parveen, and G.~Rischbieter,
  ``{Investigating the XENON1T Low-Energy Electronic Recoil Excess Using
  NEST},'' \href{http://arxiv.org/abs/2007.00528}{{\ttfamily arXiv:2007.00528
  [hep-ex]}}.

\bibitem{Baudis:2015kqa}
L.~Baudis, A.~Kish, F.~Piastra, and M.~Schumann, ``{Cosmogenic activation of
  xenon and copper},''
  \href{http://dx.doi.org/10.1140/epjc/s10052-015-3711-3}{{\em Eur. Phys. J. C}
  {\bfseries 75} no.~10, (2015) 485},
  \href{http://arxiv.org/abs/1507.03792}{{\ttfamily arXiv:1507.03792
  [astro-ph.IM]}}.

\bibitem{NDS}
M.~Verpelli and L.~Vrapcenjak, {\em LiveChart of Nuclides}.
\newblock \url{https://www-nds.iaea.org/livechart/}. IAEA, Nuclear Data
  Section, 2020.

\bibitem{betashape}
BetaShape.
  \url{http://www.lnhb.fr/rd-activities/spectrum-processing-software/}.

\bibitem{PhysRevC.91.055504}
X.~Mougeot, ``Reliability of usual assumptions in the calculation of
  $\ensuremath{\beta}$ and $\ensuremath{\nu}$ spectra,''
  \href{http://dx.doi.org/10.1103/PhysRevC.91.055504}{{\em Phys. Rev. C}
  {\bfseries 91} (May, 2015) 055504}.
  \url{https://link.aps.org/doi/10.1103/PhysRevC.91.055504}.
  
\bibitem{Aprile:2019dme}
{\bfseries XENON} Collaboration, E.~Aprile {\em et~al.}, ``{XENON1T dark matter
  data analysis: Signal and background models and statistical inference},''
  \href{http://dx.doi.org/10.1103/PhysRevD.99.112009}{{\em Phys. Rev. D}
  {\bfseries 99} no.~11, (2019) 112009},
  \href{http://arxiv.org/abs/1902.11297}{{\ttfamily arXiv:1902.11297
  [physics.ins-det]}}.  

\end{thebibliography}
\end{document}